# A Systemic Receptor Network Triggered by Human cytomegalovirus Entry

Anyou Wang[1,2*], Hong Li[1]

[1]School of Public Health, University of California, Berkeley, CA 94720,

2Human genetics, University of California, Los Angeles, CA 90095

*Corresponding Author

Anyou Wang

Email: Anyou.Wang@alumni.ucr.edu

**Abstract**

Virus entry is a multistep process that triggers a variety of cellular pathways interconnecting into a complex network, yet the molecular complexity of this network remains largely unsolved. Here, by employing systems biology approach to combine protein-interactions from existing database and genome-wide transcriptome data, we reveal a systemic virus-entry network initiated by human cytomegalovirus (HCMV), a widespread opportunistic pathogen. This network contains all known interactions and functional modules (i.e. groups of proteins) coordinately responding to HCMV entry. The number of both genes and functional modules activated in this network dramatically declines shortly, within 25 min post-infection. While modules annotated as receptor system, ion transport, and immune response are continuously activated during the entire process of HCMV entry, those for cell adhesion and skeletal movement are specifically activated during viral early attachment, and those for immune response during virus entry. HCMV entry requires a complex receptor network involving different cellular components, comprising not only cell surface receptors, but also pathway components in signal transduction, skeletal development, immune response, endocytosis, ion transport, macromolecule metabolism and chromatin remodeling. Interestingly, genes that function in chromatin remodeling are the most abundant in this receptor system, suggesting that global modulation of transcriptions is one of the most important events in HCMV entry. Results of in silico knock out further reveal that this entire receptor network is primarily controlled by multiple elements, such as EGFR (Epidermal Growth Factor) and SLC10A1 (sodium/bile acid cotransporter family, member 1). Thus, our results demonstrate that a complex systemic network, in which components coordinating efficiently in time and space contributes to virus entry. The findings and original network constructed here lay a foundation to further dissect the molecular complexity of virus entry and provide potential drug targets.

**Introduction**

Virus entry activates various cellular pathway components. For decades, intensive studies on individual genes and pathways involved in virus entry have successfully provided us with an unprecedented wealth of molecular detail on how component proteins respond to virus entry [1]. Quite unexpectedly, however, drugs targeting individual components of specialized pathways identified in single-component studies, not only failed to control HCMV infection, but also caused huge unexpected side-effects[2]. Clearly, virus entry is not simply the result of a single activated gene or pathway. Many proteins and pathways are continuously cross-talking to coordinate cellular signals during each step of virus entry, such as virus attachment, interaction with receptors, signaling, membrane fusion and endocytosis. Nevertheless, how these proteins interact with each other to permit virus entry into cells is not completely understood. In particular, very little is known about the system-wide network and functional modules involved in virus entry, what the complete virus-receptor network looks like, and which proteins control this network. This type of knowledge is the initial step towards completely elucidating the complexity of virus entry and developing efficient treatments to prevent virus spread to other cells.

HCMV is a ubiquitous opportunistic pathogen that causes fatal or permanently debilitating disease in immunologically compromised individuals and neonates. Particularly at risk for infection with this virus are AIDS patients, cancer patients, organ or tissue transplant recipients undergoing immunosuppressive therapy, infants, fetuses, and the elderly. More recently, the virus has also been implicated in tumorigenesis[3], and the etiology of circulatory diseases, most notably, atherosclerosis.

HCMV entry into cells activates (up- and down-regulates) a variety of signaling pathways and multiple cellular receptors. HCMV attachment/entry (~ 5 to ~25 min post-infection (PI) triggers components and pathways linked to receptor tyrosine kinase, mitogen-activated protein kinase signaling, cytoskeletal rearrangement, transcription factors, prostaglandins, and cytokines [4]. In particular, HCMV entry activates epidermal growth factor receptor (EGFR), $\alpha v \beta 3$ integrin ($\alpha 2\beta 1$, $\alpha 6\beta 1$, and $\alpha v\beta 3$), and their signaling pathways [5-8], which play important roles in HCMV entry. Therefore, EGFR and $\alpha v\beta 3$ integrin have been proposed as HCMV receptor and co-receptor, respectively [5-8]. However, EGFR is not expressed on all HCMV-permissive cell types that are efficiently infected by HCMV. In addition, EGFR might not be essential for HCMV entry[6], and integrins only play a role in downstream events during HCMV entry[5-8]. In vivo, HCMV can infect almost every organ system and tissue type[4,6,9], and in vitro HCMV can promiscuously penetrate diverse cell lines with varying receptors. Together, these findings suggest that HCMV entry activates multiple receptors that interact in a network, and that elucidating the HCMV entry process could provide valuable insights into the mechanism of virus entry in general.

It can be extremely challenging for traditional genetics and biochemistry to characterize a complex process like virus entry, but systems biology network approaches (e.g. system-wide protein-protein interaction network) can greatly facilitate such a task[10]. Protein interactions can be physical binding interactions (e.g. those from yeast two-hybrid) and functional interactions (e.g., via interactions with phosphate groups on phosphoproteins). These interactions can be extracted from published databases and can be consolidated into genome-wide comprehensive networks[11]. Systematic analysis of these networks can simultaneously elucidate all possible pathway components associated with viral infection and the cross-talk that occur among these components in response to virus infection.

In this study, we used a systems biology approach for the first time to systematically elucidate a comprehensive systemic network response to virus attachment/entry, reveal a comprehensive virus receptor network, and identify in silico the essential components of these networks. Our work provides a conceptual framework to further understand the fundamental molecular basis of virus entry.

**Results**

**A comprehensive protein-interaction network linked to HCMV entry**

To systematically decode the systemic network activated by HCMV entry, we first utilized systems network approaches to search published databases for all physical and functional protein-protein interactions known to date to be associated with HCMV attachment and entry (see Materials and Methods). These interactions were then combined into a systemic protein-interaction network database, which currently comprises 6651 nodes (proteins) and 64392 edges (interactions) (Figure 1A, Table S1) and is being expanded daily as databases are updated.

To examine the overall architectural features of this network, we analyzed overall node degree distribution, which represents the possibility of nodes having a given degree, and the number of incident edges to a given node. The node degree distribution of our network decreases with degree and approximates a power law (figure 1B), indicating that our network is a scale-free network, which is proposed as a universal network framework in biology networks[12-14]. In addition, we also calculated the average of the clustering coefficient C(k) distribution, which describes how nodes link to others via their K neighbor to form clusters or groups. C(k) also diminishes with the increase in number of neighbors (figure 1B), indicating that our network is a hierarchical network [12-14] predominated by hubs (highly connected proteins) and bottlenecks, which are nodes with many shortest paths going through them analogous to key bridges that link sub-networks to a whole map network [15]. Both hubs and bottlenecks likely play essential roles in this type of networks [12-15]. These distribution properties of our network are similar to other biological networks previously reported[12-14].

**HCMV entry activates a complex systemic network**

After constructing the comprehensive network database, we next enriched the network (figure 1) with genome-wide transcriptome data on genes significantly altered by HCMV attachment and entry, 5 min and 25 min respectively post infection of human primary foreskin fibroblasts, a common cell line used as a model of HCMV infection. A total of 408 and 240 genes showed significantly altered gene expression patterns at 5 min and 25 min PI respectively (Table S2-S3). These enriched networks became systemic networks activated by HCMV attachment and entry (figure. 2). A total of 7 functional modules (figure 2) were activated at 5 min PI, including phosphorylation, intercellular junction assembly, iron transport, cell differentiation, vesicle-mediated transport, immune response, chromatin disassembly, macromolecule metabolism, cell communication, and signal transduction. At 25 min PI, 3 functional modules were activated, including immune response, transmembrane receptor protein tyrosine kinase signaling pathway, and sodium ion transport. This rapid decrease in the number of activated genes from 408 (5 min PI) to 240 (25 min PI) (Table S2-S3) within a very short time interval after HCMV infection, and the decline in network modules (from 7 to 3) fit the biphasic model of cellular response to infection [4], in which activation of cellular signaling peaks immediately in response to infection, then rapidly declines dramatically, and then gradually increases again to another peak.

While modules of receptor system, ion transport and immune response dominated the entire process of HCMV entry, cell adhesion and skeletal movement were featured at 5 min PI and immune response predominated in the network at 25 min PI (figure 2). Notably, genes were not co-regulated in most of the modules activated during HCMV entry (figure 2), indicating that HCMV entry not only modulates different complex modules, but also mediates components within the same module. HCMV entry thereby elaborates module functions of the same module.

**A systemic receptor network involved in HCMV entry**

Genes that are down-regulated could play an important role in HCMV attachment and entry, but the receptor system, in particular, should be up-regulated during these stages of infection [5-7]. Since the network comprising down-regulated genes did not have any characterized functions (figure S1), we focused on a systemic receptor network containing 123 proteins (Table S4) that were enhanced at both time points (5 min and 25 min PI). This network was decomposed into 7 functional groups, including macromolecule metabolism and chromatin remodeling, signal transduction, cell surface receptor pathway, skeletal development, immune response, endocytosis, and ion transport (figure 3 ). Consistent with previous reports about HCMV entry[4,5,9], the network includes almost all known pathways and their components up-regulated by HCMV entry. Such pathways include receptor- like EGFR in the receptor group, mitogen-activated protein kinase-like MAPK10 (mitogen-activated protein kinase 10) in the signaling group, components for cytoskeletal rearrangement in the skeletal group, transcription factors located in the nucleus, cytokines located in the extracellular space, and components for calcium transport in the ion transport group. Importantly, our network also revealed a systemic view of the HCMV receptor system, in which genes are clustered into multiple functional groups of varied pathways, and simultaneously performing various functions and bioprocesses during HCMV infection. For example, the receptor pathway group contains 18 different receptor components (EGFR, TP73L, CCR5, OR1A1, TCF4, AVPR1B, RELA, GLP1R, GNAO1, SOST, ADRA1A, GNG4, DGKA, PRB4, NRP1, DOK2, SORCS2, PTPRS), the skeletal group 14 components (DLX2, BAPX1, SGCA, LMO2, HOXA2, IBSP, COL9A2, RUNX1, EGR1, ANKH, CSRP3, ANXA13, NPR3, SOX6), and the ion transport group 8 components (SLC34A2, ATP7A, TRPM1, MBP, SLC10A1, VMD2, TRPC5, SLC17A2). While signal transduction for cell communication and cell adhesion dominated the network, components for macromolecule metabolism and chromatin assembly or disassembly were surprisingly the most abundant in the network. This indicated that chromatin remodeling is one of the major bioprocesses occurring in the human host during HCMV attachment. These data reveal a complex HCMV receptor system that comprises several functional sub-networks that are functionally dominated by signal transduction, cell adhesion, and chromatin remodeling.

**Key proteins in the HCMV receptor network**

To identify the essential components in the HCMV receptor network, we examined the contribution of individual components to the network by knocking out single genes in silico. Special attention was paid to protein components located in the extracellular space and cell membrane (figure 3) because these components play crucial roles in initiating bioprocesses during HCMV entry, or serve as potential HCMV receptors. After knocking out individual genes, we calculated the alterations in the average number of neighbors, which describes the contribution of individual nodes to network connectivity, and the mean shortest path, which measures the smallest number of links between selected nodes and essentially indicates network diameter. Node knockouts in a network would decrease network connectivity. Moreover, knockout of nodes that are higher in the network hierarchy, would result in greater reduction of connectivity. As for diameter, the longer the diameter, the less interconnectivity there is in the network. Knocking out a hub would increase diameter because of the loss of short paths in a network, whereas knocking out a bottleneck would decrease diameter because the network would be broken down and the long path that normally link to sub-networks would be lost.

Results of in silico knock out experiments showed that the component EGFR contributed most in network connectivity and diameter (figure 4 A-B), indicating that it serves as a hub (highly connected proteins) in the HCMV receptor network. Similarly, IL4 (interleukin 4), KRAS (kirsten rat sarcoma 2 viral oncogene homolog), and IBSP (integrin-binding sialoprotein) also serve as hubs in this network activated by HCMV entry. In contrast, whereas CLU (clusterin) and SLC10A1 are also major contributors to network connectivity, knocking them out resulted in a decrease in network diameter (figure 4 B) (figure 4A), indicating that these two components serve as bottlenecks in this network stimulated by HCMV attachment and entry.

To confirm in silico the consequence of knocking out these hubs and bottlenecks, the structure of mutant and wild-type network activated by HCMV attachment and entry was compared (figure 5). While knocking out hubs like EGFR altered linkages of local sub-network (figure 5A and 5B), knocking out bottlenecks CLU and SLC10A1 completely broke down the entire network into at least two independent sub-networks as highlighted in figure 5C. These results indicate that both potential hubs (EGFR, IL4, KRAS, and IBSP) and bottlenecks (CLU and SLC10A1) identified above are real hubs and bottlenecks in this network, and that they play important roles in this network. This finding is similar to reports about other networks in which hubs and bottlenecks are likely more essential in a network than other nodes [12-15].

## Discussion

### HCMV entry activates a complex systemic network

Studies on virus entry using traditional genetics and biochemistry approaches, have identified several viral entry pathways into host cells. [1,5-7]. However, the molecular mechanisms underlying virus entry remain largely elusive. We systematically assembled the existing databases of all pathway components into a systemic scale-free network to elucidate the complexity of HCMV entry (figure 1). The advantage of systems network approach is that it accounts for all interactions and cross-talks among components and treats the whole interactions as a network instead of linear circuits explicated by conventional approaches. The cross-talk that has been mostly ignored in conventional studies can significantly contribute to real phenotypes[16] and they were included in the present systemic network. The network constructed in the present study is based on current database. Future database updates and system-wide protein data may slightly change the linkages in our network; moreover, our network data need to be verified by direct experimental evidences like those in systems biology approaches. However, the overall architecture of our network database is not expected to change significantly because of its stable universal features, and scale-free and hierarchical structure (Figure 1). Therefore, the network constructed in this study can be adapted to analyze molecular mechanisms of host-microbe interactions in general, and can potentially find application in drug discovery against virus entry.

Entry of infectious agents into host cells activates complex bioprocesses [1,17-19]. Previous studies demonstrated that HCMV entry stimulates gene expression of various pathway components, such as those involved in immune response, calcium transport, and signal transduction [4-7,9,20,21]. In the present study, we systematically identified a systemic network and dynamic molecular modules activated by HCMV entry, which includes, not only genes and pathways previously reported, but also those uncovered in the present study (figure 2). Network module identification could be affected by network sources and some identified modules might not be related to biological functions[22]. However, since our network interactions were weighed on the basis of source confidences, and then network topology features and gene functions from gene ontology databases were combined to identify the network modules, our set of modules in this study likely represents a featured class of protein complexes in a natural biological network stimulated by HCMV entry. These functional modules can be used to further model the quantitative contribution of signal transduction of pathway components to HCMV entry by systems biology approaches.

The dynamic functional modules identified here were activated immediately upon infection. Modules of cell adhesion and skeletal movement were activated at ~5 min PI (figure 2), suggesting that HCMV enters the cell much earlier than thought. Notably, genes in most of the modules were not co-regulated (figure 2), suggesting a greater molecular complexity of HCMV entry than previously thought.

**A complex receptor network involved in HCMV entry**

Infectious agents can easily bind to cell surfaces via chemical interactions, but with low affinity. Microbe-specific receptors and co-receptors are required to strengthen these bindings, but they are not likely sufficient for a successful entry, which require subtle contributions from other functional groups. For instance, calcium transport and cytoskeletal movement, which are often observed during microbe entry, are essential for surviving some receptor-ligand interactions and play crucial roles in strengthening microbe-attachment to cell surface [23]. Similar roles are true for signal transduction, immune response, and chromatin remodeling [23]. Therefore, a complex coordinated network is required for microbe entry into cells, but has not been elucidated until now. [1,18,19,24]. Here, our data revealed a HCMV receptor network that includes, not only receptor sub-network, but also chromatin remodeling, signal transduction, skeletal development, immune response, endocytosis, and ion transport (figure 3). Since this network contains all pathway components known to date to be related to virus entry, this network probably represents a complete coordinated network sufficient to mediate HCMV entry. Surprisingly, genes associated with chromatin remodeling were the most abundant in this HCMV receptor network, suggesting that chromatin remodeling is a major event during HCMV entry.

Bioprocesses associated with microbe entry are similar for different microbe species, but the pathway components mediating these bioprocesses are usually species-specific. Particularly, cellular receptors are highly species-dependent. As for HCMV, integrin cellular co-receptors facilitate HCMV entry, but they only work in downstream receptor-ligand interactions during HCMV entry [5,8,9]. Indeed, a successful integrin-ligand high affinity attachment depends on how molecules underneath the membrane surface respond to integrin-ligand adhesion [23]. Other proteins, such as focal adhesion kinase, phosphatidylinositol phosphate kinase, and F-actin, need to be activated before integrin receptor activation [23,25]. Over-expression of genes in the receptor and signal transduction groups (figure 3) might account for the integrin activation. For example, PIP5K3 (phosphatidylinositol-3-phosphate/phosphatidylinositol 5-kinase, type I) regulates actin cytoskeleton and focal adhesion; Dok2 (docking protein 2) plays a crucial role in integrin outside-in signaling through a physical and functional interaction with integrin $\alpha v \beta 3$; MAPK10 (mitogen-activated protein kinase 10, MAP kinase activity) plays a key role in focal adhesion; RAP2A (RAS related protein 2a) engages beta2 integrins; IBSP (integrin-binding sialoprotein) interacts with integrins for cell adhesion. These findings further argue for integrins acting as downstream co-receptors for HCMV entry; meanwhile, the identity of the HCMV receptor remains elusive.

Multiple receptors have been proposed for HCMV entry, but they have not been unambiguously identified [4-7,9]. Some of the 18 members of the receptor group in figure 3 likely act as HCMV receptors. In particular, the genes that are essential in the HCMV receptor network might be much more important for HCMV entry than receptors defined previously because of their essentiality in this network. Generally, hubs and bottlenecks are likely essential in a network [12-15]. By knocking out genes in silico, we identified EGFR, IL4, KRAS, and IBSP as hubs; and CLU and SLC10A1 as bottlenecks in the HCMV receptor network (figure 4-5). Hubs and bottlenecks are new emerging concepts, and there is no available standard algorithm to identify them, so far. Identifications of hubs and bottleneck may be biased depending on the algorithm and network resources used to construct the network. We merged all databases in our study (figure 1) to eliminate database bias, and the node contributions for both network connectivity and diameter calculated here (figure 4-5) were consistent with those for network centrality[26] that are essential for a network (data not shown). Therefore, the hubs and bottlenecks identified here are likely essential in the natural HCMV-receptor network and constitute the group of proteins that are likely essential for HCMV entry.

As a member of hubs, EGFR was previously reported as an essential component of the HCMV receptor network although this result needs to be confirmed [6,7]. More detailed attention should be paid to the annotation of *egfr* genes used in studies because, there are three annotated *egfr* genes in the human genome; namely, accession number # AF277897 (located in chr7:55,200,539-55,203,821), #U95089 (chr7:55,054,067-55,192,136), and #U48722 (chr7:55,054,221-55,192,136). Correspondingly, there are three probe-sets in the Affymetrix chip: 1565484_x_at, 210984_x_at, and 211607_x_at. In our gene expression experiments, expression of the *egfr* gene corresponding to accession #AF277897 was up-regulated, but the other two *egfr* genes were down-regulated. We focused on the EGFR with accession #AF277897 because its expression was enhanced at both time points (5 min and 25 min PI). Our network data also showed that the same EGFR plays an important, if not essential, role in HCMV attachment and entry, at least at the early stage (figure 4-5). A similar role was found for the other hubs. KRAS is a protein in the small GTPase superfamily that is activated by integrins during virus entry. KRAS also interacts with multiple immune receptors and is involved in multiple pathways related to cell adhesion and virus entry, such as regulation of actin cytoskeleton, tight junction, EGFR-ErbB (erythroblastoma viral gene product homolog) signaling pathway, and mitogen-activated protein kinase (MAPK) signaling (http://www.genome.jp/kegg/). IL4 is a cytokine that facilitates virus entry[27]. IBSP is a sialoprotein that could bind to integrin as another component in the HCMV receptor system[5,8]. Two glycoproteins (SLC10A1, CLU) were identified as bottlenecks (figure 4-5). SLC10A1 (solute carrier family 10) belongs to sodium/bile acid cotransporter family. Ion transport plays an important role in integrin binding during virus entry as discussed above. In addition, SLC10A1 is also involved in lipid and lipoprotein metabolism (http://www.reactome.org/cgi-bin/eventbrowser?DB=gk_current&ID=73923), which might be related to the lipid rafts that signal during virus entry. CLU (clusterin) is one of the sulphated glycoproteins that is activated by virus infection[28] and regulates cell communication and signal transduction related to infection like the lectin-induced complement pathway (http://www.invitrogen.com/content.cfm?pageid=10878), and the NF-kappaB pathway[29]. Thus, these hubs and bottlenecks identified here are biologically important for HCMV entry.

In this study, we used gene expression data to enrich the protein-interaction network. This activated network may not be completely consistent with those derived from protein level data, but genomics data measured by the Affymetrix microarray employed here are generally overlapping with the proteomics data [30 ]. Our findings about the complex network activated by HCMV entry and the HCMV receptor network should emphasize the molecular complexity of virus entry. Targeting one or two receptor proteins as currently employed may not efficiently block virus entry and prevent virus spread across cells. The rapid change in dynamic modules makes it challenging to develop an efficient strategy to block virus entry, but the receptor network identified here and the approach we have developed should lay a framework to further dissect the molecular complexity of virus entry and facilitate efficient drug development.

**Methods and materials**

**Network construction**

Physical interactions were extracted from protein-protein binding database ([BIND](BIND), [DIP](DIP), [HPRD](HPRD), [PreBIND](PreBIND), [http://bond.unleashedinformatics.com/Action](http://bond.unleashedinformatics.com/Action), [http://dip.doe-mbi.ucla.edu/](http://dip.doe-mbi.ucla.edu/), [http://hprd.org/](http://hprd.org/), [http://www.blueprint.org/products/prebind/index.html](http://www.blueprint.org/products/prebind/index.html),). Functional interactions were derived from a functional experimental database and literature mining, including curated inflammatory disease database, EMBL human database [11,31,32], Cancer Gene Curation database([http://ncicb.nci.nih.gov/NCICB/projects/cgdcp](http://ncicb.nci.nih.gov/NCICB/projects/cgdcp)), and database for *Agilent Literature Search* ( [http://www.labs.agilent.com/sysbio/](http://www.labs.agilent.com/sysbio/)) [31] including PubMed([http://www.ncbi.nlm.nih.gov/entrez/query.fcgi](http://www.ncbi.nlm.nih.gov/entrez/query.fcgi)), OMIM ([http://www.ncbi.nlm.nih.gov/entrez/query.fcgi?db=OMIM](http://www.ncbi.nlm.nih.gov/entrez/query.fcgi?db=OMIM)), and USPTO ([http://patft.uspto.gov](http://patft.uspto.gov)). The network was visualized by Cytoscape [33] and is shown in Figure 1. The complete network is listed in table S1.

**Virus and Cells**

HCMV Towne strain (American Type Culture Collection) was used to infect primary human foreskin fibroblasts (CC-2509) at a MOI of 10 as previously described[34].

**RNA extraction and microarray hybridization**

Infected and uninfected cells were trypsinized and collected by centrifugation. RNA was purified using the RNeasy RNA purification kit (QIAGEN Inc. Valencia, CA), followed by DNase treatment to eliminate all traces of DNA, according to the manufacturer's recommendation. GeneChip® One-Cycle Target Labeling and Control Reagents (Affymetrix, Santa Clara, CA) were used to process RNA and for hybridization following the manufacturer's protocols. Affymetrix Human Genome U133 Plus 2.0 Arrays, which contains over 47,000 transcripts that completely cover the whole human genome, were employed in this study.

**Microarray data analysis**

Algorithms [35,36] were employed to analyze microarray data, which were normalized using the invariant set method [35,36]. The array with median CEL intensity was chosen as the baseline array. The model-base (PM/MM) expression value was calculated. To minimize false positives, two steps were performed to filter the data. The first step compared samples (arrays) in a pair-wise fashion with the threshold for fold change at > 2 with 90% confidence interval, and a difference in expression values between the genes >100. The second filter criteria was set as coefficient of variation > 0.3, and the genes that were present were called at > 20%. Genes filtered from the screening criteria were taken to be the genes with significant alteration in gene expression, and used to enrich the network.

**Network analysis**

Genes with significant alteration in gene expression at time points 0 min, 5 min, 25 min PI were used to enrich the network initially constructed (figure 1 and table S1). Upon HCMV entry, the enriched network becomes an activated (up- and down regulated) network (Figures 2 and table S2). The network was analyzed by using Network Analyzer (http://med.bioinf.mpi-inf.mpg.de/netanalyzer/index.php). The activated network was weighted on the basis of confidences of interaction sources, and the weighted networks were decomposed into functional modules based on topological interconnection intensity and gene functions (http://www.geneontology.org/) [37-40]. Genes were classified according to gene ontology database (http://www.geneontology.org/) [39].

**ACKNOWLEDGMENTS**

The authors thank Rong Hai for providing technical support.

**Figure legends**

**Figure 1. A comprehensive regulatory network linked to HCMV entry**. The network was constructed by using protein-protein binding database and protein functional database (see text and Materials and Methods for details). The insert shows a zoomed portion of entire network. The colors of nodes (proteins) and edges (interactions) represent gene expression levels and edge sources, respectively. The same color strategy for nodes and edges will be used for all figures in this study unless otherwise specified. Also shown are entire network properties, including node degree distribution that approximates a power law, $P(k) \sim k^{-\gamma}$, ($\gamma = 0.95$ in our network), and C(k) distribution, average clustering coefficient that measures the tendency of nodes to form clusters[12], which decreases with the number of neighbors.

**Figure 2. Systemic networks and functional modules activated by HCMV attachment and entry.** The complete network activated by HCMV entry is listed in tables S2 and S3. Only parts of entire networks are shown for clarity. 2A, functional modules activated at 5 min PI; 2B, activated network at 25 min PI.

**Figure 3. Network up-regulated by HCMV entry.** Genes shown here were up-regulated by HCMV entry at both 5 min and 25 min PI. Genes are clustered into functional groups and color-coded. Only the primary functions for each gene are indicated. Cellular components are shown on the left side.

**Figure 4. Contribution of individual genes to properties of the network enhanced by HCMV entry.** Extracellular and membrane components of the network enhanced by HCMV entry (figure 3) were individually knocked out in silico, and the effects of such knock out were calculated. Only genes with at least two direct neighbors in the network were knocked out because genes with only one direct neighbor or without neighbors are located at the end-terminal in the network and would not significantly affect the network architecture. 4A,

Contribution of individual genes to network connectivity. 4B, Contribution of individual genes to network diameter.

**Figure 5. Samples of in silico gene knock out in the network enhanced by HCMV entry.** 5A, Entire wild-type network enhanced by HCMV entry with arrow pointing at genes to be knocked out. Blue, potential bottleneck nodes; green, potential hubs. 5B, Knocking out hubs led to decrease in local sub-network linkages as highlighted in green circle when compared to wild-type network. 5C, Knocking out bottleneck nodes broke down the entire network into at least two separated networks as highlighted in blue circles.